# Local inhomogeneities resolved by scanning probe techniques and their impact on local 2DEG formation in oxide heterostructures

M. Rose[1,2], J. Barnett[3], D. Wendland[3], F. Hensling[2], J. Boergers[2], M. Moors[2], R. Dittmann[2], T. Taubner[3], F. Gunkel[1,2]

[1] *Institute of Electronic Materials (IWE II), RWTH Aachen University, Aachen, Germany*
[2] *Peter Gruenberg Institute and JARA-FIT, Forschungszentrum Juelich GmbH, Juelich, Germany*
[3] *I. Institute of Physics (IA), RWTH Aachen University, Aachen, Germany*

Abstract:

Lateral inhomogeneities in the formation of 2-dimensional electron gases (2DEG) directly influence their electronic properties. Understanding their origin is an important factor for fundamental interpretations, as well as high quality devices. Here, we studied the local formation of the buried 2DEG at $LaAlO_3/SrTiO_3$ (LAO/STO) interfaces grown on STO (100) single crystals with partial $TiO_2$ termination, utilizing *in-situ* local conductivity atomic force microscopy (LC-AFM) and scattering-type scanning near-field optical microscopy (s-SNOM). Using substrates with different degrees of chemical surface termination, we can link the resulting interface chemistry to an inhomogeneous 2DEG formation. In conductivity maps recorded by LC-AFM, a significant lack of conductivity is observed at topographic features, indicative of a local $SrO/AlO_2$ interface stacking order, while significant local conductivity can be probed in regions showing $TiO_2/LaO$ interface stacking order. These results could be corroborated by s-SNOM, showing a similar contrast distribution in the optical signal which can be linked to the local electronic properties of the material. The results are further complimented by low-temperature conductivity measurements, which show an increasing residual resistance at 5 K with increasing portion of insulating SrO terminated areas. Therefore, we can correlate the macroscopic electrical behavior of our samples to its nanoscopic structure. Using proper parameters, 2DEG mapping can be carried out without any visible alteration of sample properties, proving LC-AFM and s-SNOM to be viable and destruction-free techniques for the identification of local 2DEG formation. Furthermore, applying LC-AFM and s-SNOM in this manner opens the exciting prospect to link macroscopic low temperature transport to its nanoscopic origin.

Oxide interface electronics are an exciting possibility to implement highly correlated physics phenomena into modern electronics with the possibility to combine and access new functionality in electronic devices.[1, 2] The LaAlO$_3$/SrTiO$_3$ (LAO/STO) system is one of the most investigated oxide heterostructures as the 2-dimensional electron gas (2DEG) at its interface holds astonishing properties, such as high mobility at low temperatures, gate-tunable conductivity, even the usually exclusive ferromagnetism and superconductivity. All of these properties are accessible in one confined electronic system.[3-6] The 2DEG forms by charge transfer, avoiding the polarity mismatch induced by LAO deposition on TiO$_2$-terminated STO. Residing on the STO side,[7] the resulting properties of the 2DEG are strongly dependent on the balance between ionic and electronic charge transfer, which both compensate the polarity mismatch.[8] This balance is determined by prior treatment of the STO substrate, subsequent thin film deposition parameters[5, 7, 9-14] and post-deposition treatment.[15-19] Another aspect is the stacking order of the atomic planes at the interface, which is determined by the termination of the STO substrate. In LAO/STO, the interfacial stacking order of TiO$_2$/LaO gives rise to a negative diverging potential, leading to electron transfer into the STO side of the interface (also called n-type). The predicted counterpart caused by a SrO/AlO$_2$ stacking order (p-type interface),[20] is still a debated topic. Even though recent results show a hole gas can be achieved,[21, 22] usually p-type interfaces turn out to be insulating.[3, 23-25] This behavior is commonly explained by a preferential ionic compensation of the potential by incorporation of oxygen vacancies.[26-28] As a result, even enhanced oxygen ion conductivity was proposed for this interface.[29] Vertically to these interfaces, the electronic to ionic charge transfer and carrier distribution has been investigated in great detail at present, however, in the lateral dimension it is less intensely studied.

To improve the understanding of the lateral 2DEG distribution, local probing of the interface properties becomes a valuable tool. Until now, the 2DEG is typically understood as completely homogeneous on the lateral scale. Only at low temperatures, domain structures could be observed due to the STO phase transition to tetragonal crystal symmetry.[30-32] In this study, we are using local conductivity atomic force microscopy (LC-AFM) complemented by scattering-type scanning near-field optical microscopy (s-SNOM), to test this assumption. A strong electrical transport anisotropy of the 2DEG could already be observed when the STO substrate surface holds an ordered, mixed termination.[33] Further works on the self-patterning of STO surfaces confirmed the mixed termination and highlight the ability to use this property to structure interfaces.[34, 35] However, investigation of these naturally occurring inhomogeneities has not received much attention, even though research in this direction could potentially hold answers to open questions in the field: The above-mentioned lateral homogeneity of the 2DEG, causes of insulating interfacial areas, causes for sample-to-sample variation, and the role of an inhomogeneous 2DEG distribution for low-temperature phenomena. Utilizing a non-destructive technique, which allows subsequent investigation of samples by other methods to correlate macroscopic to local properties, opens the way to an enhanced understanding and functionality. Scanning probe techniques are proven to extract detailed information across sample surfaces, exceeding the mere mapping of topography. It could be shown that phase differences in tapping AFM and lateral-force differences in contact AFM are able to distinguish chemical surface terminations.[35, 36] Furthermore, s-SNOM has recently been shown to be able to optically detect free charge carriers in STO[37] as well as in LAO/STO heterostructures.[38-40] Both these established techniques (Phase contrast AFM and s-SNOM) will be used to show how LC-AFM can contribute to scientific advance in the field. Further scanning probe examples are scanning superconducting quantum interference device measurements, which revealed magnetic patches in LAO/STO that depend on the LAO thickness[41] and at low temperatures were able to identify highly conducting pathways, and strain-tunable magnetic ordering. Both phenomena correlate to the tetragonal STO domain walls, forming below 105 K.[31, 32] Other techniques like Kelvin probe force microscopy showed the influence of oxygen vacancies on the band alignment of LAO/STO[14] as well as STO[42] and magnetic force microscopy showed the dependence of magnetic moments in LAO/STO on the electron density and thickness of the heterostructure.[43] Regarding LC-AFM, it could be shown that conducting areas can be written into LAO/STO structures of subcritical thickness,[44-46] leading to the use of this effect as nano-patterning technique.[47] This in turn enabled researchers to easily fabricate one-dimensional systems to study exotic transport phenomena.[48] LC-AFM was further used to determine the 2DEG confinement in cross sectional measurements[7] and although it was also shown that LC-AFM can be used to write conducting lines, it was not used to map naturally occurring differences of local conductivity distributions, and was not demonstrated with lateral resolution on the nanoscale.

We show in this study that LC-AFM is a suitable technique to extract this local information of the LAO/STO 2DEG even though it is a buried structure. The lateral extraction of local information is on the scale of tens of nanometers and is obtained in a non-destructive manner, i.e. allowing the local mapping of conductivity without influencing interface properties. The high influence of the STO substrate surface termination on the interface conductivity is used to show how this local probing technique can directly correlate electronic transport properties to topographic features. In this way the interplay between STO topography and interfacial transport can be analyzed, leading to further insight of local 2DEG formation distribution in oxide heterostructures. Using s-SNOM, we furthermore corroborate that the lack of conductivity observed at SrO-terminated step terraces can be related to a diminished 2DEG, rather than a local thickness variation of the LAO overlayer. Therefore, the origin of macroscopic transport differences in identically fabricated heterostructures down to low-temperature conductivity can be linked to the nanoscopic origin of the 2DEG distribution.

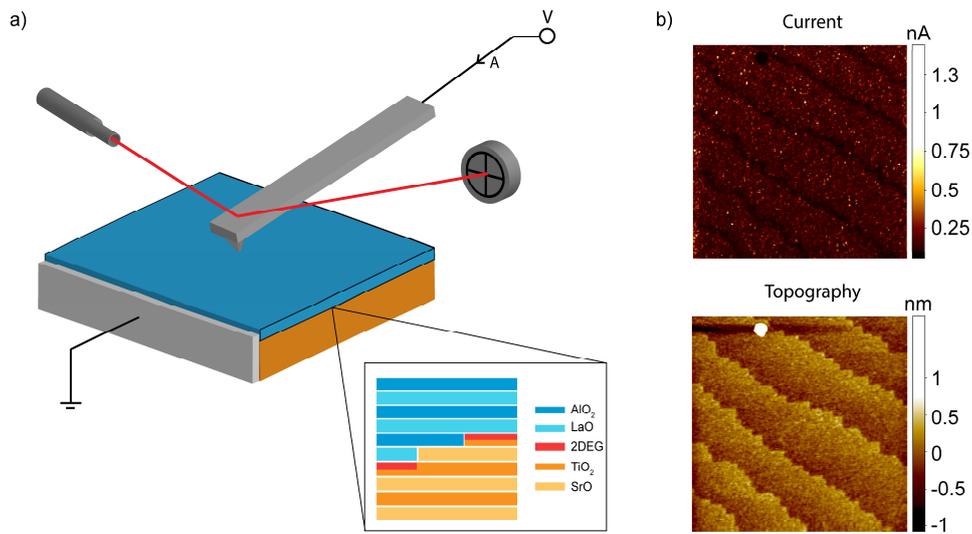

Figure 1: a) Shows a sketch of the LC-AFM system with an inset depicting the atomic structure of the samples under test; b) Shows examples of measured topography and current maps. Both images are 1x1 µm in size.

A sketch of the experimental setup is displayed in Figure 1 a. The working principle is that of classical AFM with an applied side electrode to access the laterally formed 2DEG. A voltage is applied between the doped diamond-coated AFM tip and bottom electrode while recording the current, making a parallel mapping of conductivity and topography possible. LC-AFM measurements were performed in a cluster tool, where samples can be transferred *in-situ* from the deposition chamber to the analytical methods. As samples are not exposed to air this way, the effects of adsorbates stemming from surface contaminations due to e.g. humidity can be excluded.[49-52] An example of the current and topographic contrast is given in Figure 1 b. In order to probe local 2DEG formation, we specifically selected STO substrates which show distinct discrepancies from the characteristic $TiO_2$ topography as is shown in Figure 1 b.[53] To achieve this ordered, mixed termination, we make use of a bottom-up approach, exploiting the naturally occurring variance of single crystalline STO. As-received STO single crystalline substrates typically have a mixed surface termination after they have been cut, polished, and annealed. The fabrication of $TiO_2$-terminated STO via different routes has been intensively studied using chemical and heat treatments.[35, 54-57] One common route is a wet etching treatment where water is applied to the surface, forming strontium hydroxide. A consecutive etching step in buffered hydrofluoric acid removes the hydroxide, leads to a fully $TiO_2$ terminated surface.[54] Consecutive annealing leads to the formation of an evenly distributed terrace-step structure when temperature, time and atmosphere are set correctly.[53, 58-60]

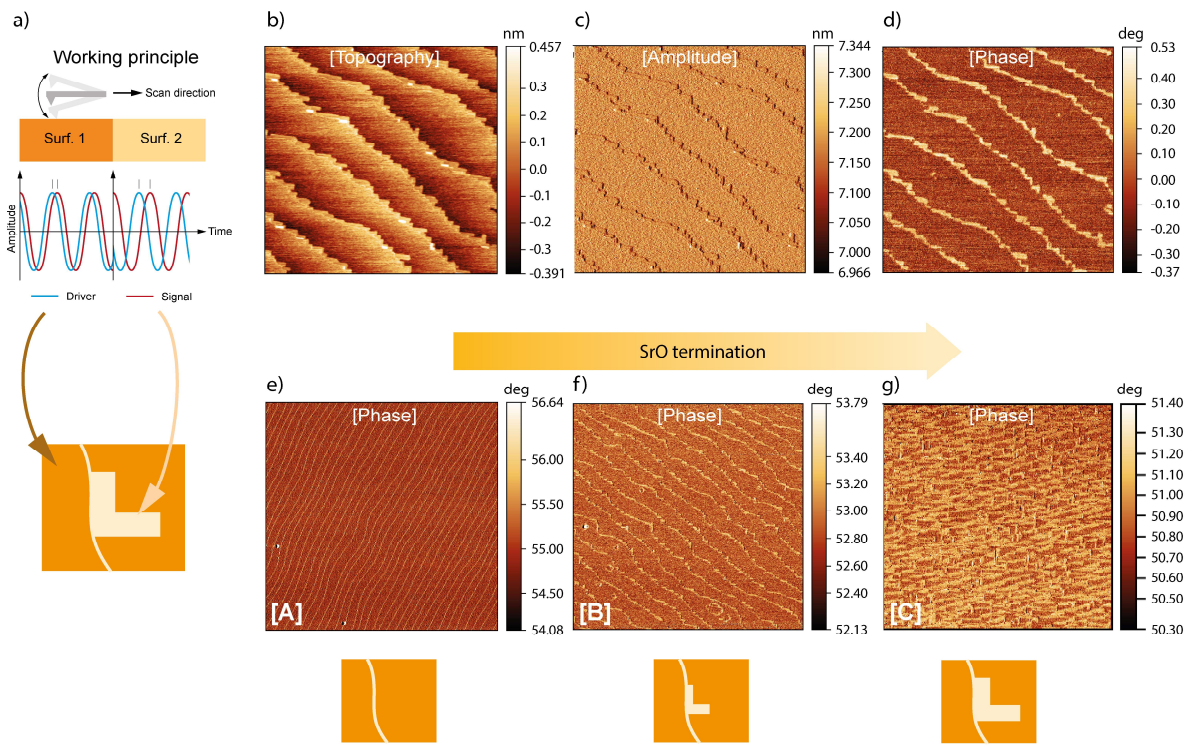

Figure 2: STO Substrates are shown with different termination distributions. All were etched with buffered HF acid and consecutively annealed at 950°C for 2h; a) shows a sketch of the working principle of phase contrast in tapping AFM; b) Topography c) amplitude and d) phase of the same measurement are shown for a sample with two different surface terminations. These images (b-d) are 1x1 µm in size; Phases are shown for samples with e) full $TiO_2$ termination (case [A]), f) partial secondary termination (case [B]), g) mixed termination (case [C]). These images (e-g) are 5x5 µm in size.

However, the properties of the as-received STO single crystals such as miscut angle [55, 61-63] and defect concentration,[64] have high impact on the surface reconstruction, influencing the final result of combined chemical and thermal treatment. For this reason, the same treatment can commonly lead to different termination ratios, when applying the same process on different STO crystals.[35, 65] The single crystalline STO(100) substrates used here, were etched in buffered hydrofluoric acid for 2:30 minutes and subsequently annealed for 2 hours at 950°C. Other routes, such as microwave induced hydrothermal treatments are alternatives to this etching procedure but require special equipment.[62, 66]

The presence of two different surface terminations was identified using phase contrast in ambient AFM measurements. In Figure 2 a, the working principle of such a phase contrast under constant amplitude imaging is shown. As the tip-to-sample interaction force is depending on the surface chemistry, the phase difference between driver and measured signal changes, when the surface termination varies locally. An example of this behavior is summarized in Figures 2 b-d. The topography image (Figure 2 b) shows rectangular topographic features present at the terrace edges in a 90° angle to each other. No roughened surface that would influence the phase contrast is detectable and the amplitude signal (Figure 2 c) stays constant except for sharp lines when the tip reaches the edge of topographic height differences. Still, the phase image (Figure 2 d) shows a strong and sharply resolved contrast that extends from the edges of the terrace structure. Hence, the rectangular features can be identified as surfaces of different local chemistry than the rest of the sample surface. Three distinct cases were observed, which will be referred to as cases [A], [B], and [C] in the further discussion and are shown in Figure 2 e-g. An example of a case [A] substrate, which represents a complete $TiO_2$ termination, is shown in Figure 2 e. This case is discernable through evenly distributed, smooth terraces. Case [B] is shown in Figure 2 f. These substrates turned out to have kinked terrace edges that differ from the ideal case [A]. As was shown in Figure 2 d, these straight features indicate a termination change present at the terrace edges, relating to a SrO termination in the case of STO.[67] Case [C] shows more pronounced features as is evident by the more extended regions in phase contrast shown in in Figure 2 g. They reach out from the terrace edges and extend over the terrace planes, in some cases even across multiple terrace steps, having a higher coverage of SrO termination than case [B]. As the interfacial transport properties of LAO/STO heterostructures depend on substrate termination,[3] it can be expected that such SrO-terminated regions should be insulating. Therefore, *in-situ* LC-AFM is applied on LAO/STO samples of mixed substrate termination to open the possibility to map the 2DEG by monitoring the local conductivity.

On each of these substrates, 4 unit cells (uc) of LAO were deposited using pulsed laser deposition (PLD) (see experimental section for details), monitored by high pressure reflective high-energy electron diffraction (RHEED). Distinct and clear RHEED intensity oscillations were observed throughout the entire thin film growth indicating layer-by-layer growth mode for all samples. A detailed characterization of thin film growth can be found in the supporting information (SI) 1. A layer thickness of 4 uc resembles the minimum thickness that leads to the formation of a 2DEG[4] and was chosen to minimize the required voltages for a detectable current through the LAO layer in subsequent LC-AFM measurements. Test samples with subcritical LAO thickness of 3 uc were fabricated and electrically characterized to ensure the absence of underlying bulk conductivity. Furthermore, measurable current can only be detected if an electrical contact between 2DEG and bottom electrode is assured. This was achieved by applying a side contact, the removal of which again results in a lack of measurable current as shown in SI 2. This experiment proves that current is transferred through the 2DEG to the ground electrode, showcasing that interfacial conductivity is accessed using this setup.

The results of LC-AFM analysis of 4 uc LAO deposited on case [A], [B] and [C] STO substrates can be found in Figure 3 a, b and c respectively. The topography of the deposited LAO thin films retains the same structural features present on the underlying substrates, as is expected for LAO thin films of this thickness. The applied voltage was set to 4 V as lower voltages lead to insufficient currents and higher voltages influence thin film morphology and sample properties.[42, 68-70] The current maps show a distinct lateral contrast that is only visible for samples that were grown on [B] and [C] case substrates where a partial SrO termination resides at the interface (Figure 3 b and c). For case [A] samples a homogeneous current map was observed as shown in Figure 3 a. The detected contrast in case [B] and [C] directly correlates with structures that differ from the smooth terrace structure of case [A] substrates. The kinked edges of case [B] substrates appear as dark region in the current maps and the same behavior is observable for the more pronounced features of case [C] substrates. Note that the contrast is neither influenced by prolonged scan times nor by the scanning direction, indicating no sample alteration or measurement artefacts through the induced current. A tunneling mechanism is most likely responsible for current in these structures, as the recorded $I(V)$ curves show a symmetric current distribution (see SI 3) and behave similar to expected $I(V)$ curves in MOS structures.[71, 72] In macroscopic tunnel junction measurements, direct tunneling was observed by Singh-Bhalla *et al.*[73] for the lower voltage regime, which transitioned into Fowler-Nordheim tunneling for the higher voltage regime when similar LAO thicknesses were used. Analyzing the $I(V)$ characteristic (See SI 3), revealed a similar behavior for low voltages, however, judging from the $I(V)$ slope at higher voltages, the most likely conduction mechanism is Poole-Frenkel emission (trap assisted tunneling) for the used setup and samples. The role of trap assisted tunneling was also determined by Swartz *et al.*[74] to have a high impact on the electrical transport through LAO thin films. This suggests that a convolution of direct tunneling and trap-assisted tunneling mechanisms is necessary to describe electrical transport in these experiments. In general, the $I(V)$ is comparable with literature and no degradation or alteration can be observed under these continuously applied voltages. Therefore, a breakdown is avoided if the voltage is not exceeding 4 V. As current can only be achieved when the 2DEG is contacted (via metallic side electrode, cf. SI 2) and no bulk conductivity is present for these samples, the lack of current is caused by the absence of interfacial conductivity. Therefore, the LC-AFM measurements shown in Figure 3 confirm the nature of interfaces grown on the different substrate cases, as mixed conducting and insulating interfaces. The SrO terminated regions determined from phase contrast AFM, lead to a SrO/LaO stacking order, as can be seen by the rectangular features corresponding to the locally insulating parts of the samples.[75] Therefore, these results confirm to first degree the insulating behavior of SrO/AlO$_2$ interfaces and the possibility to have a resulting mixed conducting and insulating interface.

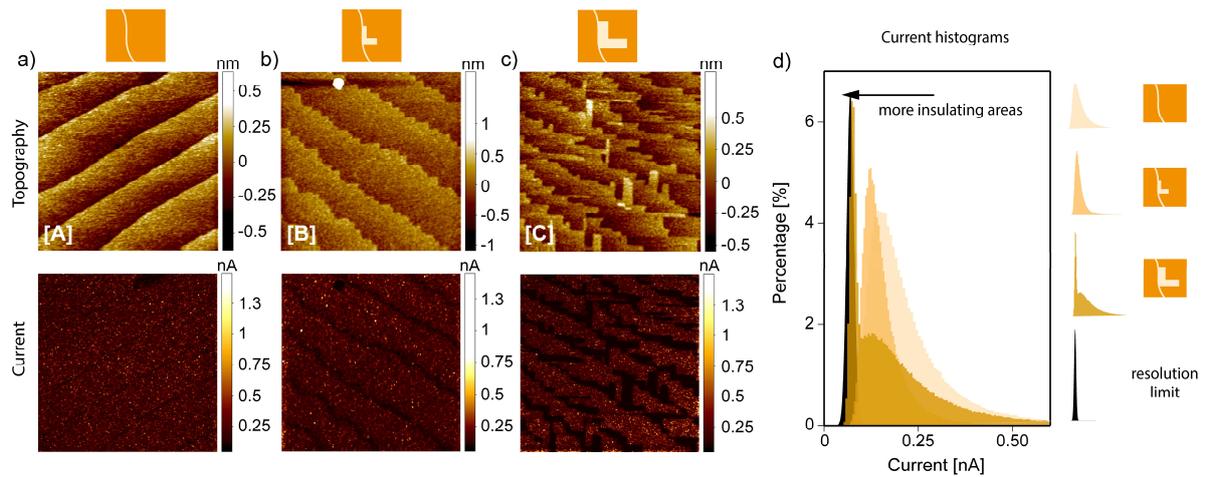

Figure 3: Topography and current maps of samples consisting of 4 uc LAO on STO with different SrO to $TiO_2$ ratios measured by LC-AFM; a) full $TiO_2$ termination; b) partial SrO termination c) mixed termination. d) shows the current map histograms corresponding to the recorded images in a)-c). All images are 1x1 µm in size.

To give a more quantified evaluation of these current maps, the corresponding histograms are displayed in Figure 3 d, showing the distribution of measured currents. A reference sample without side electrode was added to visualize the resolution limit of the used setup (black area). The case [A] sample has the maximum of its current distribution set at the highest currents, which is expected because of the evenly distributed 2DEG. The case [B] sample shows similar behavior where the maximum is shifted to lower currents as one would expect for a sample with lower conductivity. The case [C] sample shows a strikingly different distribution of current. This results from the extended dark regions that are close to the measurement limit of our setup and are visible in the histogram as a peak around the noise level of $I \approx 100$ pA. Despite the overall low currents, none of them are below the resolution limit and thus extended dark regions are effectively insulating.

To further analyze if the source of electronically induced contrast is induced via a diminished carrier density, samples that did show contrast in LC-AFM were analyzed using s-SNOM. This scanning probe technique allows the determination of local optical properties with nanoscale resolution and high surface sensitivity.[76-78] The working principle is based on a laser beam focused on a metal-coated AFM tip (Figure 4 a), leading to strong near-fields which interact via dipole-dipole coupling with the sample. Back-scattered light is demodulated at higher harmonics $n\Omega$ of the tapping frequency $\Omega$ and contains information about the dielectric function of the sample surface[79] while enabling probing depths of up to ~100 nm.[80-83] In the past, s-SNOM was shown to detect the presence of the 2DEG in LAO/STO[39] and even allow for determination of electronic properties (like carrier density and mobility) using spectroscopy at different wavelengths[38, 40] Here, a correlative s-SNOM measurement on a case [C] sample is presented at a laser wavelength of $\lambda = 10.3$ µm ($\nu = 974$ cm$^{-1}$). A distinct pattern of bright

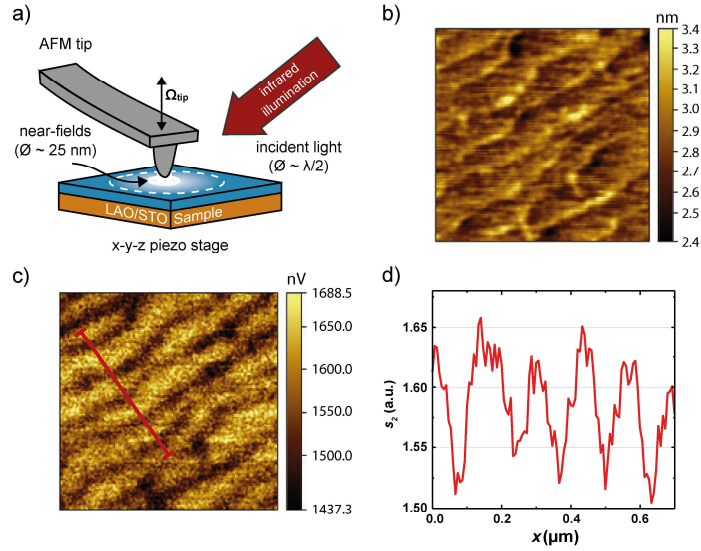

Figure 4: a) A sketch of the s-SNOM working principle is shown. b) Shows the topography image of a case [C] sample with the corresponding optical image shown in c). Both images are 1x1 μm in size. From the same measurements, single line scans of the optical signal are shown for the second order demodulation in d)

stripes and dark lines is visible in the second harmonic demodulation order (demodulation at $n\Omega$ with $n=2$) of the scattering amplitude $s_n$ (Figure 4 c). The lateral resolution in s-SNOM is broadened compared to LC-AFM due to the larger AFM tip radii introduced by metal coating. Still, the atomic step terrace structure is resolved. Topographical artifacts can be excluded from the s-SNOM signal due to the small step heights of ~0.4 nm and since no correlative behavior of the cantilever tapping phase or amplitude could be detected. The pattern in the optical amplitude (Figure 4 c) coincides with the atomic step edges visible in the topography (Figure 4 b) and similar to the conductivity maps of LC-AFM measurements (Figure 3 c) it is significantly broader than the step edges. These features show a contrast of 10% for $s_2$ and 25% for $s_3$ (line scans in Figure 4 d), which coincides well with previously reported contrast between regions with and without 2DEG of 10-30% in $s_3$ in this spectral range.[38, 39] A simulation of the s-SNOM response to varied carrier densities and mobilities showed that the amplitude of the optical signal depends more strongly on carrier density $n$ than mobility $\mu$ (see SI 4 for details of the simulation). Therefore, the dark lines in Figure 4 c can be explained as regions of a diminished 2DEG, corroborating the interpretation of LC-AFM. The combined results from phase contrast AFM, LC-AFM and s-SNOM congruently show the distinct influence of partial SrO-terminated surfaces, leading to a local 2DEG formation.

To link the LC-AFM and s-SNOM measurements to macroscopic transport, we measured the low temperature resistance for all three cases. For this purpose, sample pieces of all three cases were cut into a mimicked Hall bar geometry, where the terrace edges are aligned diagonal to the applied current. The results are summarized in Figure 5. The resistance of the case [A] sample decreases with a decreasing temperature and reaches a constant value of about 200 Ω at 5 K, showing a metallic behavior. A similar behavior can be observed for the case [B] sample but shifted to overall higher resistances. In this case the metal-like behavior is observable with a residual resistance of about 400 Ω at 5 K. Again, the case [C] sample shows distinctly different behavior. At higher temperatures it overlaps with the data obtained for case [B], but with decreasing temperature it deviates in a resistance upturn and settles at about 2000 Ω at 5 K, showing a Kondo-like behavior.[5, 84] As the conducting landscape is interrupted by the insulating areas, the resistance upturn to lower temperatures could be related to a freeze out of percolation pathways as the temperature is decreased (note that the dielectric constant of STO is strongly temperature-dependent).[85] Therefore, the macroscopic conductivity is influenced by the nanoscopic structures measurable in LC-AFM, as the highest resistance at low temperature correlates to the largest amount of insulating areas detected by LC-AFM. Furthermore, from low-temperature Hall measurements, it can be extracted that not only the average carrier density is lowered but also significantly the mobility of charge carriers (see SI 5 for additional data). From this point, it can be concluded that additional scattering mechanisms are introduced, which are most likely linked to the increased number of insulating areas the electrons have to overcome, when passing through the sample. The correlation between insulating areas and resistance shows that the electrical conductivity from room- to low-temperature is significantly reduced by the introduction of local inhomogeneities. These inhomogeneities could only be identified through application of LC-AFM, giving a simple method to identify relevant features that are only tens of nanometers large. Additionally, by s-SNOM it can be confirmed that a lack of charge carriers at the topographic features correspond to a SrO termination. The local inhomogeneities and their distribution throughout the 2DEG is directly influencing macroscopic transport behavior not only by lowering the conductance but giving rise to distinctly altered temperature behavior. Overall, the results are consistent with charge transfer being dependent on the local interface chemistry, leading to insulating behavior for SrO termination and metallic behavior only for $TiO_2$ termination.

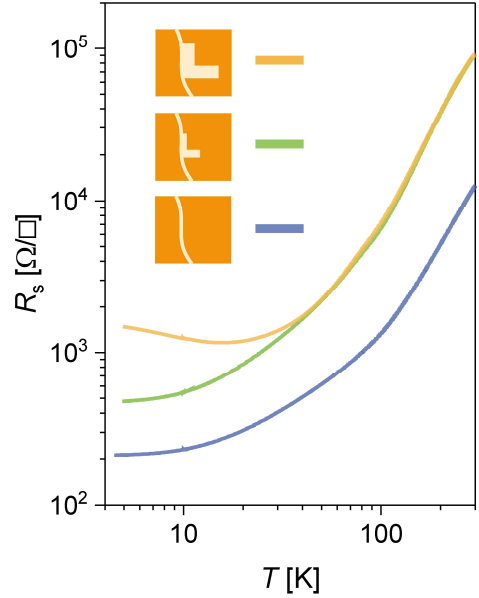

Figure 5: Low-temperature resistivity measurements of a case [A], [B] and [C] sample

**Conclusion**

We have demonstrated, for the example of the 2DEG model heterostructure LAO/STO, that in-situ LC-AFM can be used to not only access the buried 2DEG, but also to resolve inhomogeneities on the nanometer scale. We have used this concept here to locally resolve atomic terrace phenomena at the LAO/STO interface. These terrace phenomena could be linked to a variation in termination that leads to a locally distributed 2DEG formation. The results show that one can access the buried two-dimensional conductivity with this technique, enabling a lateral resolution of features that are in the range of tens of nm but can be theoretically reduced to a few nanometers. Correlative s-SNOM measurements lead to the same conclusion, confirming the presented interpretation of the results. In the future, s-SNOM spectroscopy of the 2DEG[40] could be used to quantify the local charge carrier density for inhomogeneities. In particular, simulations show that changes to the charge carrier density and mobility could yield a different spectroscopic response in this sample (cf. SI 4). The presence of insulating as well as conducting areas on a nanoscale level result in significant alterations of the low-temperature behavior, emphasizing the impact of local inhomogeneities on low-temperature transport behavior. The conductivity in these structures is not only decreased by the incorporation of insulating areas but altered into an upturn of resistance to the lowest temperature for the more pronounced cases. Through the combined methods, the extended structural terrace features could be identified to act as scattering centers for electrons. This enables a destruction-free, nanoscopic and simple method to identify naturally occurring inhomogeneities on the lateral scale, which can be easily combined into the overall characterization of buried 2DEG systems. These results show how nano-scaled changes in a system have significant influence on the macroscopic scale and set a baseline for further studies towards the identification of causes for the natural inhomogeneities in local 2DEG formation.

**Experimental**

**PLD Process:**

All samples were grown in the same PLD chamber (Surface Systems and Technology GmbH & Co. KG, Germany) under the same nominal conditions. Thin films were ablated from a single crystalline LAO target that was placed at a distance of 54 mm from the STO substrates. The substrates were glued on sample holders using silver paste and heated up using an infrared laser heater system to 800 °C. For ablation, an excimer laser (248 nm) with a repetition rate of 1 Hz and a fluence of 0.96 J/cm$^2$ was used. The deposition pressure inside the chamber was set to 1x10$^{-3}$ mbar. After growth, samples were left at deposition conditions for 1 hour to refill growth-induced oxygen vacancies.[7, 10]

***In-situ* LC-AFM:**

Samples were transferred *in-situ* into the LC-AFM chamber using a cluster system. Here samples can be transferred a pumped tunnel with a low base pressure of below 1×10$^{-8}$ mbar. The LC-AFM is part of an Omicron VT-SPM system (Omicron Nano Technology GmbH) and was operated using a NANOSENSORS™ CDT-FMR tip. The pressure inside the LC-AFM system was always at 1×10$^{-9}$ mbar during measurement. The tip was scanned with a speed of 1 µm/s under a force of 2 nN over the respective surfaces. For the shown measurements, a voltage of ± 4 V was applied.

**AFM:**

AFM measurements in ambient conditions were done using a Cypher AFM system from Asylum Research in tapping mode. The surfaces were scanned under a constant amplitude with a scanning speed of 5 µm/s.

**s-SNOM:**

The s-SNOM measurement was done using a commercial microscope (neaSNOM, Neaspec GmH) with a pseudoheterodyne module, a mercury cadmium telluride detector (InfraRed associates), and illumination at $\nu$ = 974 cm$^{-1}$ from a continuous wave $CO_2$ laser (Edinburgh Instruments) with a high signal-to-noise ratio. The topography signal and the near-field optical amplitude are recorded simultaneously and the optical amplitude $s_n$ is demodulated at higher harmonics $n\Omega$ of the tip oscillation frequency $\Omega \approx$ 250 kHz, with $n$ = 2 and 3, to suppress background contributions. Commercially available AFM tips were used as scattering tips, in this case Pt-Ir-coated silicon tips with a tip radius of $r$ < 30 nm (ArrowNCPt, NanoAndMore), and the tapping amplitude was adjusted to approximately $A$ = 35 nm.

**Low-temperature resistivity:**

The resistivity at low temperatures was measured using a PPMS system from Quantum Design. Sample contacts were achieved by wire bonding Aluminum directly onto the sample surface. Samples were cooled down to 5 K while measuring the resistance in Hall bar geometry.

**Conflicts of interest**

The authors declare no conflicts of interest.

**Acknowledgements**

This work was supported by the German Research Foundation (DFG). F. Gunkel and M.-A. Rose acknowledge financial support from under Grant Agreement DFG FG 1604 (No. 315025796). J. Barnett and T. Taubner acknowledge the financial support under Grant Agreement no. TA 848/7-1.

# Supporting Information

# Local inhomogeneities resolved by scanning probe techniques and their impact on 2DEG formation in oxide heterostructures

M. Rose[1,2], J. Barnett[3], D. Wendland[3], F. Hensling[2], J. Boergers[2], M. Moors[2], R. Dittmann[2], T. Taubner[3], F. Gunkel[1,2]

[1] Institute of Electronic Materials (IWE II), RWTH Aachen University, Aachen, Germany
[2] Peter Gruenberg Institute and JARA-FIT, Forschungszentrum Juelich GmbH, Juelich, Germany
[3] I. Institute of Physics (IA), RWTH Aachen University, Aachen, Germany

## S1: Pulsed Laser Deposition

All LAO thin films shown in the original manuscript were grown in the same PLD chamber under the same conditions (see experimental details). In Figure S1 a summary of thin film growth of a 4 uc LAO/STO sample on a substrate of mixed termination is shown. The intensity of the primary reflected RHEED spot is plotted over time in Figure S1 a and shows clear oscillations from beginning to end of the deposition. The substrate showed a typical RHEED pattern that is shown in Figure S1 b that is still visible after deposition with an overall lowered intensity (Figure S1 c). The topography of the substrate, which can be seen in Figure S1 d, is qualitatively replicated by the LAO thin film (Figure S1 e). All in all the deposition of the LAO thin films on substrates of mixed termination does not show discrepancies to the ones deposited on perfectly terminated STO.

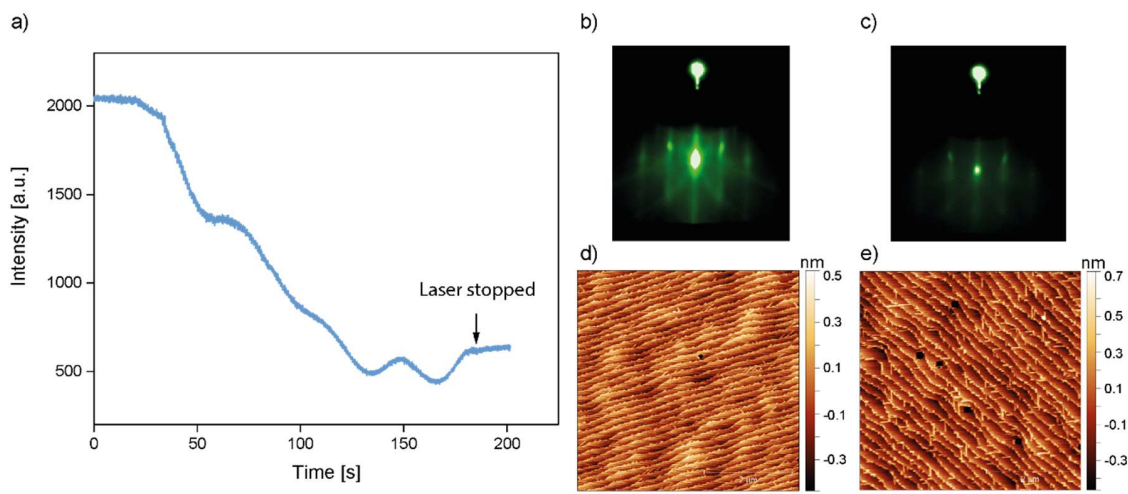

*Figure S1: A summary of the LAO thin film growth is shown a) plots RHEED intensity of the primary reflected spot over time b) shows the RHEED pattern before deposition and c) after deposition d) shows the AFM measurement of the STO substrate and e) of the LAO thin film*

## S2: Proof of principle

In this experiment the ability of the LC-AFM setup to detect currents through the 2DEG of the LAO/STO sample was tested. For this purpose, a sample was first measured with the LC-AFM without side contact. The result can be seen in Figure S2 a, which shows the recorded topography in the top image and the local current map in the bottom image. No currents above noise level could be measured in this configuration. In the second step, a side electrode using silver paste was applied to the sample and the measurement was repeated. Note that after such *ex-situ* steps, the sample was heated inside a preparation chamber at 200 °C for 30 min to desorb any surface contaminations that may be present due to air exposure. The result is shown in Figure S2 b, where now a significant enhancement of measurable current is visible. Removing the side electrode in the third step (Figure S3 c) leads to the vanishing of measurable currents. Through this series of experiments it can be deduced that current is indeed coupled into the 2DEG of the LAO/STO sample and therefore depends in its magnitude on its local electrical resistance.

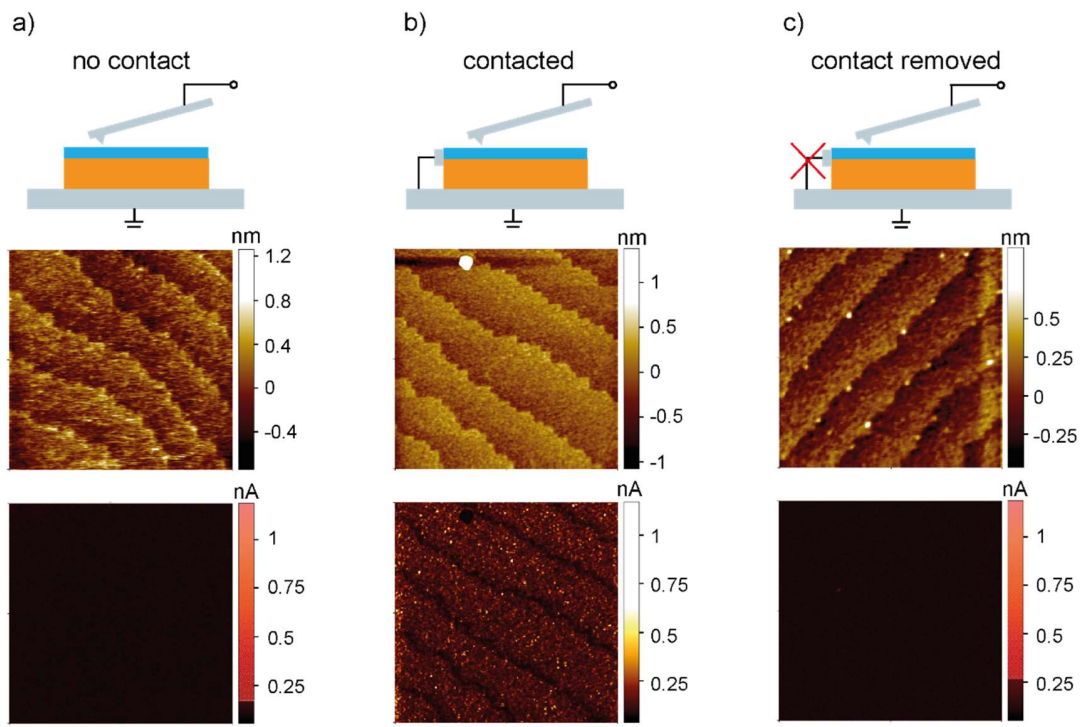

*Figure S2: A consecutive series of LC AFM measurements are shown with a) the sample contacted at the backside of the substrate b) the sample contacted through a side electrode c) after removal of the side electrode*

**S3: Current voltage characteristic of LAO/STO samples in LC-AFM**

Here the characteristic behavior of current to voltage ($I(V)$) for the measured currents of a 4 uc LAO on STO sample is shown in the used LC-AFM system (Figure S3). For this the tip was held at one position while the voltage amplitude was swept 10 times from -4 to +4 $V$ and back. The curve corrected to the offset of about 100 pA that is present in this setup (also in vacuum without a sample). In these measurements it can be seen that aside from statistical variance, the sample does not show any change in its $I(V)$ behavior over the whole experiment. Furthermore it can be shown that significant currents can be detected that do not lead to an alteration of sample properties. The $I(V)$ curve shows a symmetric behavior hinting at a tunneling mechanism. To investigate the nature of the tunneling mechanism, the characteristic plots for direct tunneling ($J \propto V$), Fowler-Nordheim ($\ln(J/V^2) \propto 1/V$) and Poole-Frenkel emssion ($\ln(J) \propto V^{1/2}$) are shown in Figure S4. For this analysis, first the average of the I(V) voltage sweeps was determined to minimize the noise level of the measurement. The result is shown in standard linear form in Figure S4a and in logarithmic form in Figure S4b. The noise level is sufficiently suppressed by the averaging. The characteristic plot for Fowler-Nordheim emission is shown in Figure S4c in which no clear linearity can be seen over the used voltage range, leading to the conclusion that it is not the dominant contribution to the $I(V)$ curve. The characteristic plot for Poole-Frenkel emission is shown in Figure 4d, where to high voltages a linearity is visible to high voltages. For high voltages the I(V) characteristic can be therefore described by a Poole-Frenkel emission which requires the existence of trap states inside the LAO thin film. This transport mechanism was already proposed by Swart et al. [1] for the transport across Co/LaAlO$_3$/SrTiO$_3$ heterojunctions. The complete I(V) curve, however, cannot be described by one single transport mechanism. Most probably a convolution of multiple conductions is taking place, making the exact determination more complicated. For a better understanding of the exact transport phenomena, temperature dependent measurements would be needed. In summary a convoluted transport is observed that can be described by a trap assisted tunneling process at high voltages.

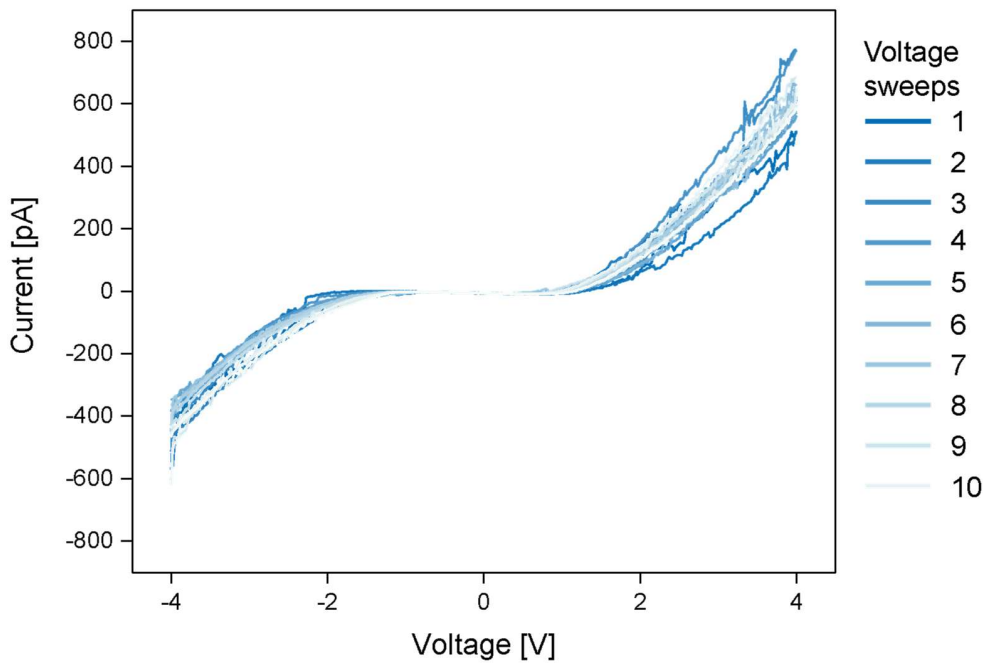

*Figure S3: The measured current over applied voltage is shown for a sample of 4 uc LAO on STO in the LC-AFM setup used. The colors from deep to light blue each represent a consecutive voltage sweep.*

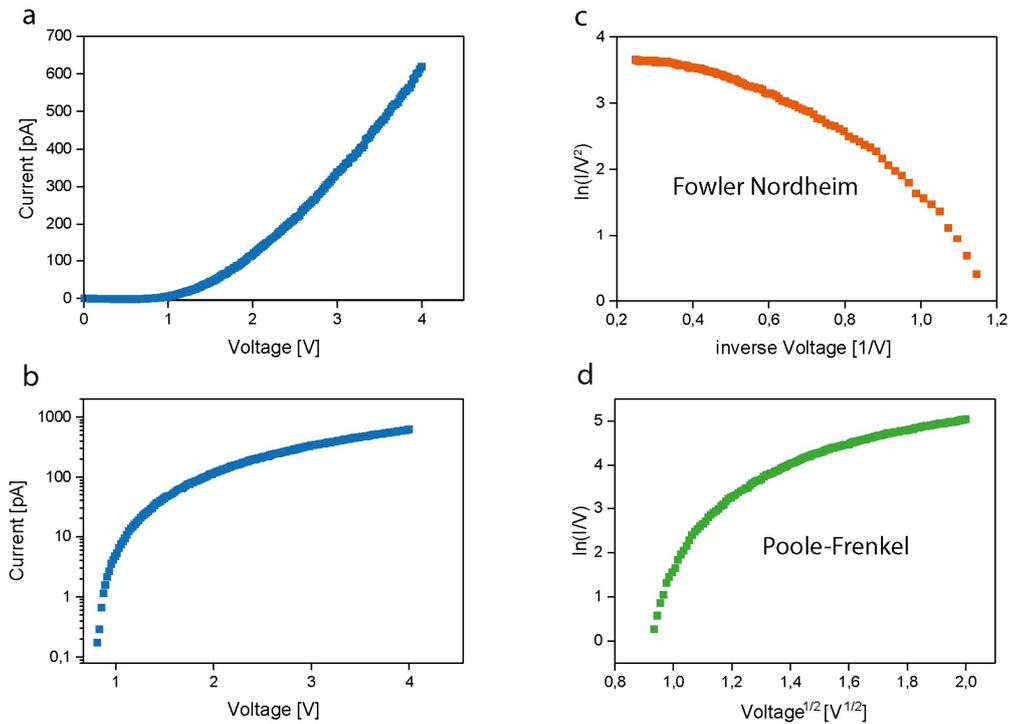

Figure S4: a) shows the current average resulting of ten voltage sweeps, b) shows a) on a logarithmic scale, c) shows the ln(I/V$^2$) over 1/V that is characteristic for Fowler–Nordheim emission d) shows ln(I/V) over the square root of V which is characteristic for Poole-Frenkel emission

## S4: SNOM contrast dependence on electronic properties

To model the s-SNOM contrast, the Finite Dipole Model (FDM) [2] was employed in combination with a Transfer Matrix Method (TMM), resulting in a powerful predictive tool [3] to simulate the near-field response of arbitrarily layered systems with known dielectric function. SNOM contrast simulations of the LAO/STO 2DEG were already published [4] and are utilized here with averaged values obtained from Hall measurements: $n_{2D}$ = 9.96×10$^{13}$ cm$^{-2}$, $\mu$ = 4.09 cm²/Vs. In order to obtain a volume concentration, the sheet carrier concentration is distributed over a thickness of $d_{2DEG}$ = 7.5 nm, resulting in a volume concentration of $n_{3D}$ = 1.33×10$^{20}$ cm$^{-3}$. To present a realistic frame of reference, we take into account non-averaging behaviour due to the insulating step edges and the fact that the 2DEG might be confined to a smaller thickness, yielding higher charge carrier concentrations. As a result, we look at a variation between $n_{3D}$ = 1×10$^{20}$ cm$^{-3}$ and 6×10$^{20}$ cm$^{-3}$ (cf. Figure S6e). Similar bounds can be established for the mobility, which is expected to stay below $\mu$ = 10 cm²/Vs at room temperature. Finally, it is important to note that the 2DEG is located within the STO side and is therefore modeled as free electrons (Drude contribution) on a background given by STO phonons (Lorentz oscillator). As such, the dielectric function of the 2DEG layer is identical to that of the STO substrate, as long as no free charge carriers are present.

Figure S6 shows the SNOM contrast when adding a 2DEG of (volume) charge carrier concentration $n$ to a formerly insulating LAO/STO sample, starting with changes to the dielectric function of pure STO (black) in Figures S6a (real part) and S6b (imaginary part). Here, a rising $n$ is marked by a transition to red color (cf. Figure S6e,f for explicit values of $n$). Using these dielectric functions for the 2DEG layer, the SNOM contrast of LAO/STO samples is presented in Figures S6c (amplitude) and S6d (phase), indicating changes to both quantities at the measurement frequency of $\nu$ = 974 cm$^{-1}$ (vertical grey line). In particular, a shifting flank is visible, that is related to the zero-crossing of Re[$\varepsilon$], which moves to higher frequencies for rising $n$. Finally, Figures S6e and S6f show SNOM amplitude and phase as a function of $n$. Here, the near-field amplitude in Figure S6e shows a strong increase of up to 350% (compared to $n$ = 0) for the upper bound of $n$.

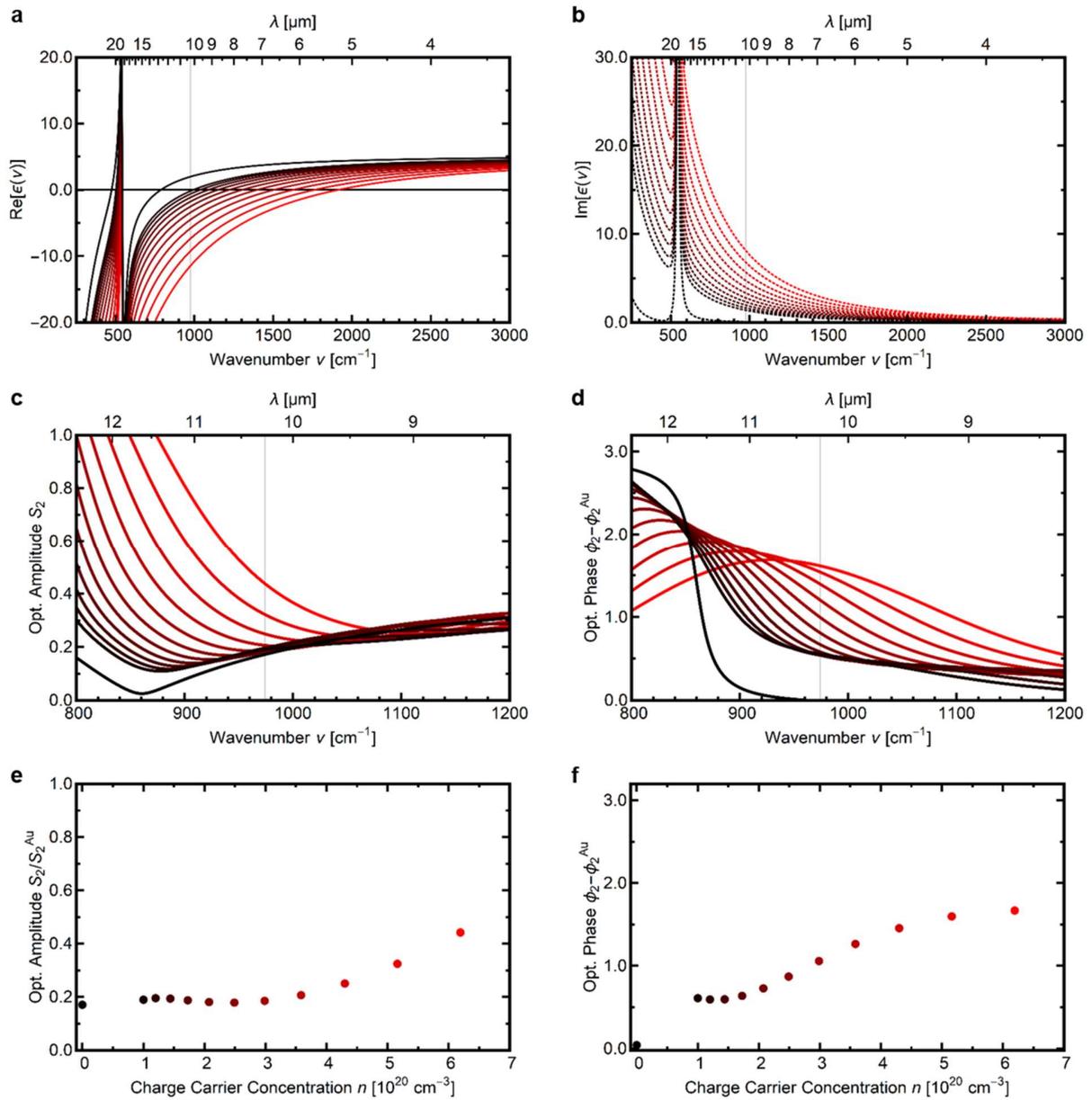

*Figure S5: Simulation of SNOM contrast dependence on charge carrier concentration: **a** real part of the dielectric function $Re[\varepsilon]$, showing the phonon resonance of pure STO (black) and the addition of a 2DEG with rising charge carrier concentration n (transition to red, values shown in e and f). The zero-crossing shifts to higher frequencies with rising n; the vertical grey line indicates the measurement frequency of $\nu = 974$ cm$^{-1}$. **b** corresponding imaginary part $Im[\varepsilon]$, showing an overall increase with rising n. **c** Simulation of normalized SNOM amplitude and **d** phase resulting from changes to the dielectric function with rising n. **e,f** changes to the SNOM-contrast as a function of n at the measurement frequency $\nu = 974$ cm$^{-1}$.*

Figure S7 shows a variation of SNOM contrast with charge carrier mobility, to compare the influence of the two electronic parameters $n$ and $\mu$. Comparing Subfigure S6e and S7e, the influence of low $n$ and $\mu$ values on the near-field amplitude is similar, but whereas a higher $n$ would lead to increased amplitude contrast, it would decrease for higher $\mu$ for this sample.

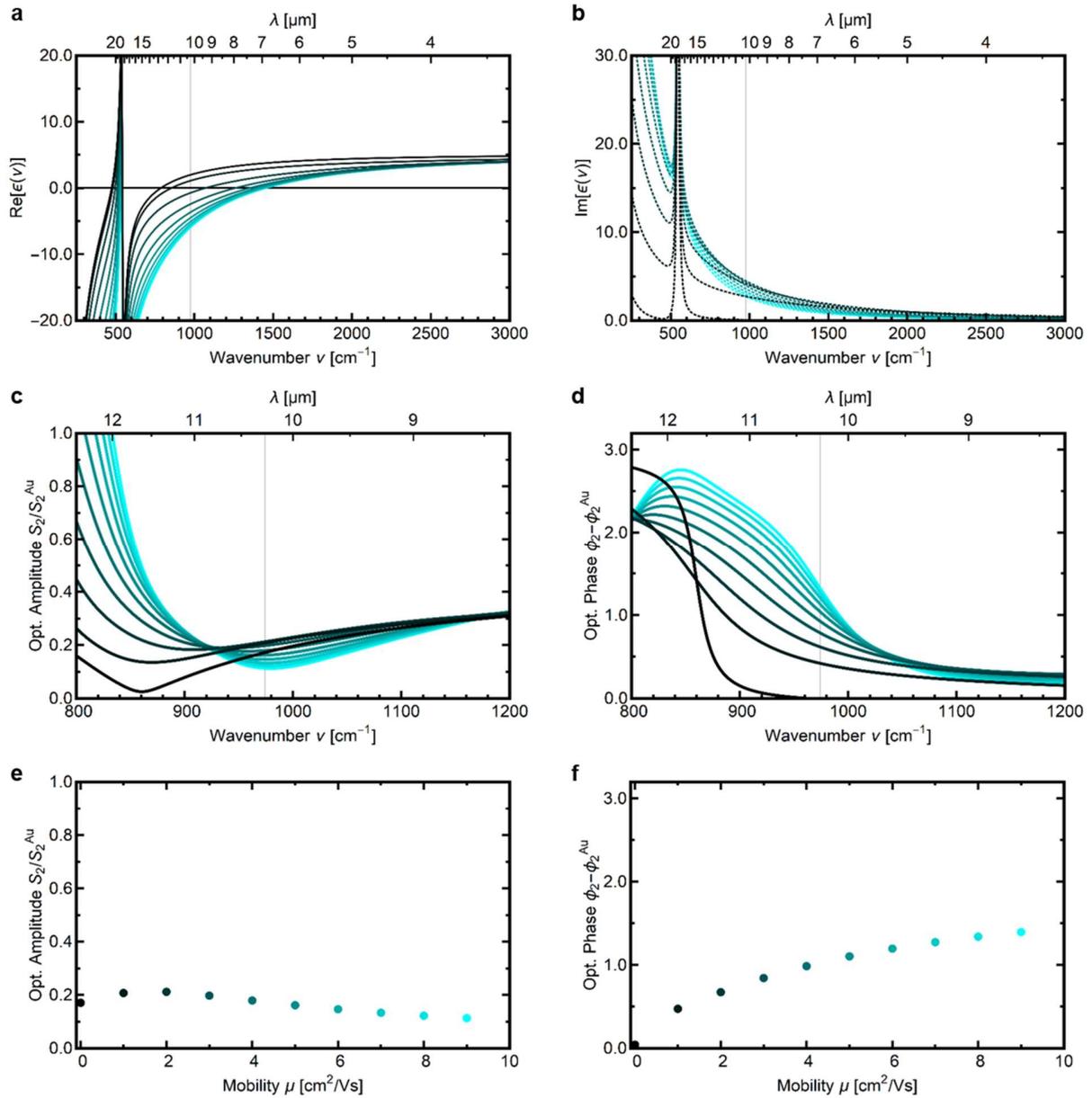

*Figure S6: Simulation of SNOM contrast dependence on charge carrier mobility; similar to figure S6. **a, b** changes to the dielectric function of the 2DEG layer (cf. **e, f** for explicit values of the color scale). **c, d** resulting near-field spectra in amplitude and phase. **e, f** contrast as a function of charge carrier mobility.*

## S5: Low-Temperature Hall analysis

Here the Hall measurements conducted in low temperatures are shown in Figure S4 for a case [A] and [C] sample. The data was recorded using a mimicked Hall bar geometry where contacts were achieved by Al wire bonding. All shown data was furthermore symmetrized to compensate offset voltages[5]. In the temperature regime below 100 K the linear Hall effect observed in

LAO/STO at room temperature becomes non-linear. This behavior results from the two available conduction bands that cannot be distinguished at higher temperatures due to phonon scattering. The carrier populations ($n_1$, $n_2$) and mobilities ($\mu_1$, $\mu_2$) can be extracted from the Hall coefficient ($R_H$) under an applied magnetic field ($B$) by the use of a two carrier model that follows the expression of [6]

$$R_H = -\frac{1}{e} \frac{\left(\frac{n_1 \mu_1^2}{1+\mu_1^2 B^2} + \frac{n_2 \mu_2^2}{1+\mu_2^2 B^2}\right)}{\left(\frac{n_1 \mu_1^2}{1+\mu_1^2 B^2} + \frac{n_2 \mu_2^2}{1+\mu_2^2 B^2}\right)^2 + \left(\frac{n_1 \mu_1^2}{1+\mu_1^2 B^2} + \frac{n_2 \mu_2^2}{1+\mu_2^2 B^2}\right)^2 B^2}$$

In Figure S4 a) the Hall coefficients ($R_H$) of a case [A] (blue) and [C] (orange) sample are shown over the applied magnetic field at a temperature of 5 K. The fit of $R_H$ (red) is only applied in the range between 3 and 9 T as in the middle region an anomalous Hall effect additionally alters the shape of the two carrier model. It is more strongly pronounced for the case [C] sample than for case [A] but the higher overall resistance leads to additional noise on the measurement, making a clear conclusion more difficult. The anomalous Hall effect is generally an indication of magnetic moments which suggests that the LAO/STO heterostructures of mixed termination show a stronger magnetic signature. However, as the situation at these interfaces is rather complicated with multiple percolating 2DEGs, further investigations are needed to make this claim. In Figure S4 b) the extracted carrier densities are shown for the applied model over temperature. Both carrier populations are lower for the case [C] sample in respect to case [A]. The one of higher density ($n_1$), corresponding to a low mobility ($\mu_1$ in Figure S4 c), by only a small amount while the carrier population of lower density ($n_2$), corresponding to a high mobility ($\mu_2$ in Figure S4 c), by almost an order of magnitude. Qualitatively this can already be seen by the overall higher magnitude of $R_H$ in case [C]. Also both mobilities are lower in the case [C] sample, however both carrier types seem to be lower by roughly the same factor in this case. In summary both carrier densities as well as their mobilities are lowered when SrO is present at the LAO/STO interface.

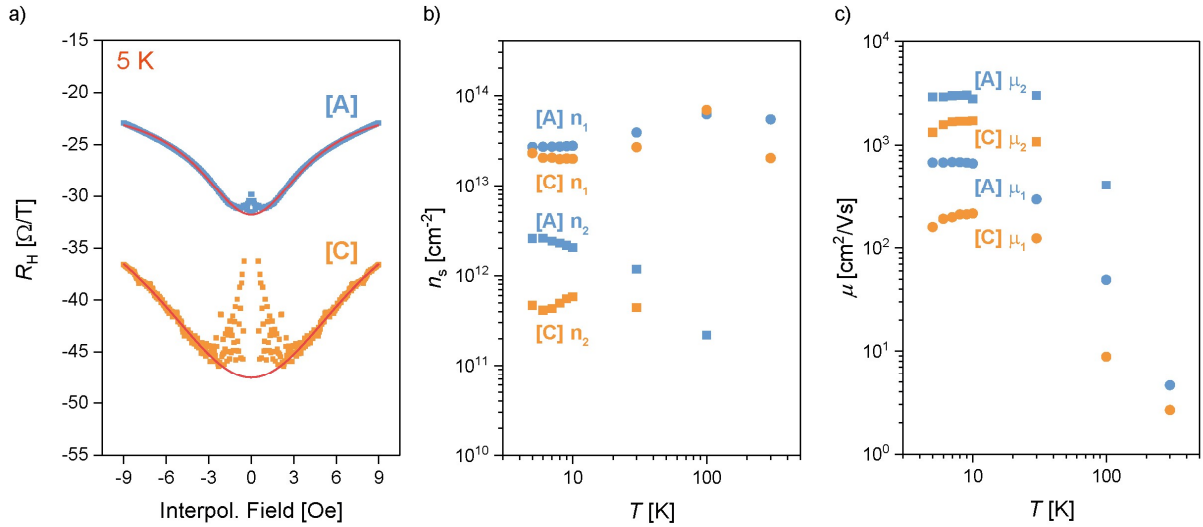

*Figure S7: a) The Hall resistances $R_{xy}$ over applied magnetic field for a case [A] (blue) and [C] sample (orange) are shown at a temperature of 5 K. b) The extracted carrier densities $n_1$ (dots) and $n_2$ (squares) over temperature are shown for a case [A] (blue) and [C] sample (orange). c) The extracted mobilities $\mu_1$ (dots) and $\mu_2$ (squares) over temperature are shown for a case [A] (blue) and [C] sample (orange).*